\documentstyle[twoside,fleqn,espcrc2,epsfig]{article}

\newcommand{\AmS}{{\protect\the\textfont2
  A\kern-.1667em\lower.5ex\hbox{M}\kern-.125emS}}

\font\sfhuge= cmss24  at 20truept
\font\sfLARGE= cmss14  at 14truept
  at 12truept
\font\sfmed= cmss10  at 10truept
\font\sfsml= cmss8   at  8truept

\title{Event Shapes from JADE Data and Studies of Power Corrections}

\author{P.A.~Movilla~Fern\'andez\address{III. Physikalisches Institut A,
        RWTH Aachen --- Physikzentrum, \\ 
        Sommerfeldstr., D-52056 Aachen, Germany}%
       }
       
\begin{document}

\addtolength{\textheight}{-68mm}
\begin{titlepage}
\thispagestyle{empty}

\vspace*{-10mm}
\vbox to 245mm{

\hbox to \textwidth{ \hsize=\textwidth
\vbox{
\hbox{
\epsfig{file=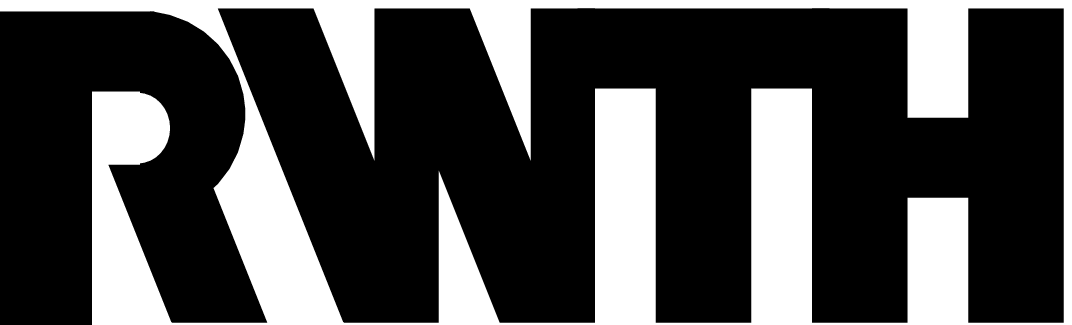,height=20mm}
}
}
\vbox{
{
\hbox{\sfmed RHEINISCH-\hss}\vspace*{+0.150mm}
\hbox{\sfmed WESTF\"ALISCHE-\hss}\vspace*{+0.150mm}
\hbox{\sfmed TECHNISCHE-\hss}\vspace*{+0.150mm}
\hbox{\sfmed HOCHSCHULE-\hss}\vspace*{+0.150mm}
\hbox{\sfmed AACHEN\hss}\vspace*{0.200mm}
}
}
\vbox{ \hsize=58mm 
{
\hspace*{0pt\hfill}\hbox{\sfLARGE\hspace*{0pt\hfill}        PITHA 98/24\hss}\vspace*{-2mm}
\hspace*{0pt\hfill}\hbox{        \hspace*{0pt\hfill} \rule{45mm}{1.0mm}\hss}
\hspace*{0pt\hfill}\hbox{\sfLARGE\hspace*{0pt\hfill}       August 1998\hss}\vspace*{2.3mm}
}
}
}

\vspace*{5cm}

\begin{center}
{\huge\bf  Event Shapes from JADE Data \\[1mm] and  \\[4mm] Studies of Power Corrections}

\end{center}
\vspace*{2cm}
\begin{center}
\Large
Pedro~A.~Movilla~Fern\'andez \\
\bigskip 
\bigskip
III. Physikalisches Institut, Technische Hochschule Aachen\\
D-52056 Aachen, Germany
\end{center}

\vspace*{0pt\vfill}
\vfill

\vspace*{-5mm}
\noindent
\hspace*{-5mm}
\hbox {
\rule{\textwidth}{0.3mm}
}

\vspace*{3mm}
\noindent
\begin{minipage}{\textwidth}

\vbox {\vsize=60mm
\hbox to \textwidth{\hsize=\wd0
\hbox {\hspace*{-5mm}

\vbox{ 
\hbox to \textwidth{\hss\sfhuge PHYSIKALISCHE INSTITUTE\hss }\vspace*{2.0mm}
\hbox to \textwidth{\hss\sfhuge      RWTH AACHEN\hss }\vspace*{2.0mm}
\hbox to \textwidth{\hss\sfhuge D-52056 AACHEN, GERMANY\hss}
}

}
}
}

\end{minipage}
}

\end{titlepage}

\setlength{\textheight}{202mm}
\setlength{\textwidth}{160mm}
\setlength{\oddsidemargin}{-4mm}
\setlength{\evensidemargin}{4mm}
\setlength{\topmargin}{16mm}
\setlength{\headheight}{13mm}
\setlength{\headsep}{21pt}
\setlength{\footskip}{30pt}
\clearpage

\begin{abstract}
  \vspace*{-52mm} \hbox{\sfsml Talk presented at the {\em QCD'98
      Euroconference}, Montpellier, France, July 2-8, 1998.  }
  \vspace*{48mm} 
  Studies of event shape observables at PETRA and LEP energies are
  presented. Previously published determinations of the strong coupling
  constant $\alpha_s$ at $\sqrt{s}$ = 35 and 44~GeV are complemented
  using new resummed QCD calculations for the $C$ parameter and
  improved calculations for the jet broadening variables $B_T$ and
  $B_W$ which recently became available.  Furthermore, recently
  predicted power corrections to the differential distributions of
  these observables are investigated. In this study, e$^+$e$^-$ data
  between $\sqrt{s}$ = 35 and 183~GeV are considered.
\end{abstract}

\maketitle

\section{INTRODUCTION}
Data from the intermediate energy region of e$^+$e$^-$ annihilation,
as provided by the JADE experiment at PETRA, allow significant tests
of perturbative and non-perturbative aspects of Quantum Chromodynamics
(QCD). This was stressed in a recently published analysis of JADE
data~\cite{bib-fernandez} where determinations of the strong coupling
constant $\alpha_s$ at $\sqrt{s}$ = 22 to 44~GeV were presented, which
are the most precise ones at PETRA energies.  Since the analysis is
based on event shape observables for which resummed QCD
predictions~\cite{bib-NLLA} (NLLA) were developed long after the PETRA
shutdown, the theoretical uncertainties of previous
measurements~\cite{bib-bethke} could be reduced significantly. By
applying LEP-established techniques of estimating experimental and
theoretical systematics, the running of $\alpha_s$ was tested {\em
  consistently} within a large range of energy
scales~\cite{bib-fernandez}.

A reliable estimation of the higher order uncertainties of the
predictions, individually for each observable, is supported by
considering as many observables as possible.  Hence, an update of
determinations of $\alpha_s$ at PETRA energies~\cite{bib-fernandez-2}
is performed, now additionally considering new calculations for the
$C$ parameter~\cite{bib-catani}.  Furthermore, improved calculations
for the jet broadening variables $B_T$ and $B_W$
\cite{bib-dokshitzer-br} are applied.

JADE data also provide clearer evidence for non-perturbative effects
\cite{bib-fernandez,bib-biebel}, which typically contribute with
reciprocal powers of $\sqrt{s}$.  They are accounted for by so-called
{\em power corrections} which, on general grounds, are found from
several theoretical approaches such as renormalons
\cite{bib-renormalon} or the dispersive approach
\cite{bib-dokshitzer}. In the following, new predictions for power
corrections to the differential distributions are tested using JADE
data and also LEP data up to $\sqrt{s}$ = 183~GeV.

\section{\label{sec-as_update} EVENT SHAPES AND $\mathbf{\alpha_s}$}

The JADE detector, one of the experiments at the PETRA e$^+$e$^-$
collider, was designed as a hybrid 4$\pi$-detector to measure charged
and neutral particles. It was operated from 1979 until 1986 at
centre-of-mass energies of $\sqrt{s}$ = 12.0 to 46.7 GeV.  This
analysis focuses on 20926 events at 35~GeV and 6158 events at 44~GeV
passing the standard JADE multihadron selection~\cite{bib-fernandez}.

\subsection{Event Shapes}
The $\alpha_s$ analysis is based on the event shape distributions of
the thrust $T$, heavy jet mass $M_H$, the differential 2-jet rate
using the Durham jet finder, $D_2$, the total and wide jet broadening
$B_T$ and $B_W$ and the $C$ parameter (cf.
\cite{bib-fernandez,bib-fernandez-2}). For the latter, no experimental
results have previously been presented at PETRA energies.

The data are corrected for detector effects and initial state photon
radiation using a parton shower plus string fragmention model (JETSET,
cf. \cite{bib-fernandez}) and a full detector simulation.  As pointed
out in \cite{bib-fernandez,bib-fernandez-2}, there is good agreement
of the Monte Carlo model with the data, despite the larger
hadronization effects at lower energies.

\subsection{Update of $\alpha_s$}
The strong coupling constant $\alpha_s$ is determined by $\chi^2$ fits
of combined NLLA~\cite{bib-NLLA,bib-catani,bib-dokshitzer-br} and
${\cal O}$($\alpha_s^2$)~\cite{bib-ERT} calculations to the event
shape distributions after unfolding for
hadronization~\cite{bib-fernandez-2}. The previously published
measurements of $\alpha_s$ at 35 and
44~GeV~\cite{bib-fernandez} are complemented by using new resummed
calculations for the $C$ parameter~\cite{bib-catani} and by applying
improved resummed calculations for $B_T$ and
$B_W$~\cite{bib-dokshitzer-br}. In the latter case, all single
logarithmic terms of the prediction are now handled more accurately
by a proper treatment of the quark recoil against multi-gluon
emission.

In the case of the $C$ parameter, the combined resummed plus fixed order
theory gives an excellent description of the data over the entire
range of the distribution, even at the low PETRA
energies~\cite{bib-fernandez-2}. The recoil-corrected calculations on
the jet broadenings do not yield a significant improvement of the
previous fits~\cite{bib-fernandez,bib-fernandez-2}. The excess of the
theoretical curve over the data in the 3-jet region is still present,
but the $\alpha_s$ results are now shifted by about 3\% to higher
values.

\begin{figure}[!t]
\vspace*{-3mm}
\centerline{
\epsfig{file=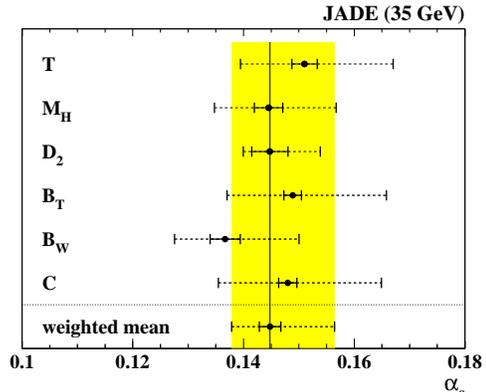,width=.45\textwidth,clip=}
}
\vspace*{-11mm}
\caption{\label{fig-as} \small
  Values of $\alpha_s$(35~GeV) obtained from ${\cal
    O}(\alpha_s^2)$+NLLA fits to the event shapes. The experimental
  and statistical uncertainties are represented by the solid, the
  total error by the dashed error bars.}  \vspace*{-5mm}
\end{figure}

The values of $\alpha_s$ and the errors obtained at $\sqrt{s}$~=
35~GeV are shown in Fig.~\ref{fig-as}, indicating that the results are
consistent with each other.  The combination of the individual results
of the six observables using the weighted average
method~\cite{bib-fernandez-2} yields as final results
\begin{eqnarray*} 
\alpha_s(35{\mathrm{GeV}}) = 0.1448 \pm 0.0010{\mathrm{(stat.)}}
  ^{+0.0117}
  _{-0.0070}{\mathrm{(syst.)}} \\
\alpha_s(44{\mathrm{GeV}})= 0.1392 \pm
  0.0017{\mathrm{(stat.)}}  ^{+0.0106}
  _{-0.0073}{\mathrm{(syst.)}}
\end{eqnarray*}
Various uncertainties emerging from experimental measurement, event
selection, hadronization corrections and the choice of the
QCD-renormalization scale were regarded as systematic errors of the
fits~\cite{bib-fernandez-2}.

\section{POWER CORRECTIONS}

\subsection{Parametrization}
A technique for studying power suppressed contributions to event
shapes is the dispersive
approach~\cite{bib-dokshitzer,bib-dokshitzer-3}.  A recent
publication~\cite{bib-dokshitzer-3} points out that the effect of
power corrections on the event shapes mentioned above is simply a {\em
  shift} $\delta V$ of the perturbative spectrum by an amount
proportional to $1/Q$ and $\ln Q/Q$, where $Q$ = $\sqrt{s}$.  The
shift depends on two parameters ${\cal P}$ and ${\cal P^{\prime}}$ and
an observable-dependent coefficient $c_V$:
\begin{displaymath}
  \delta V =  c_V \cdot \left\{ 
     \begin{array}{ll} \hspace*{-2mm}
    {\cal P} & \hspace*{-3mm} ; V = T,\, C \\ \hspace*{-2mm} 
    \left({\cal P}\ln \frac{Q}{\mu_I}- (\xi{\cal P}+{\cal P^{\prime}})\right) &\hspace*{-3mm} ;V = B_T,\, B_W.
      \end{array} \right.
\end{displaymath}
${\cal P}$ and ${\cal P^{\prime}}$ can be expressed in terms of
effective coupling moments $\alpha_0$ and
$\alpha_0^{\prime}$~\cite{bib-dokshitzer-3},%
{\small%
\begin{eqnarray*}%
{\cal P}=  a\frac{\mu_I}{Q} \hspace*{-1mm} \left[ \alpha_0(\mu_I)-\alpha_s(Q)
    -\beta_0\frac{\alpha_s^2}{2\pi}\left(\hspace*{-.5mm} \ln \frac{Q}{\mu_I}+\frac{K}{\beta_0}+1
     \hspace*{-.5mm}\right)  \hspace*{-1mm} \right] \\
{\cal P}^{\prime}=  a \frac{\mu_I}{Q} \hspace*{-1mm} \left[ \alpha_0^{\prime}(\mu_I)+\alpha_s(Q)
    +\beta_0\frac{\alpha_s^2}{2\pi} \left(\hspace*{-.5mm} \ln \frac{Q}{\mu_I}+\frac{K}{\beta_0}+2
     \hspace*{-.5mm} \right) \hspace*{-1mm} \right]
\end{eqnarray*}%
}%
where $a = \frac{4C_F}{\pi^2}{\cal M}$ contains the so-called Milan
factor ${\cal M}$ arising from a two-loop calculation of the
observable-dependent factor.  The Milan factor is expected to be
universal~\cite{bib-milan}.  $\mu_I$ is an infrared matching
scale~\cite{bib-dokshitzer}, usually set to 2~GeV.  An important
feature of the dispersive approach is that the only free parameters
$\alpha_0$ and $\alpha_0^{\prime}$ are expected to be approximately
{\em universal}~\cite{bib-dokshitzer,bib-dokshitzer-3}.

\begin{figure*}[!t]
  \vspace*{-10mm}
  \centerline{\hspace*{0mm}\epsfig{file=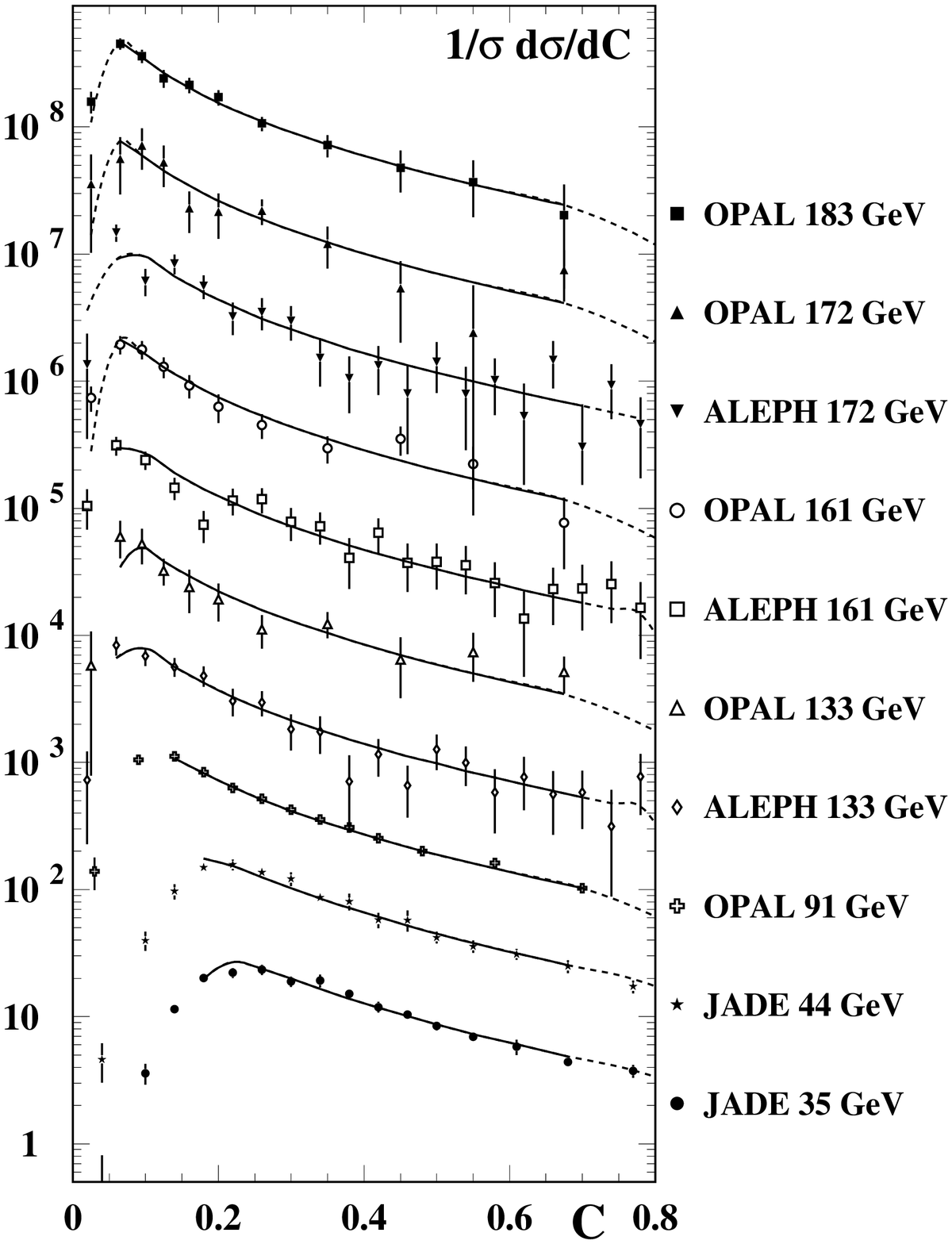,width=64mm}
    \hspace*{-15mm}\epsfig{file=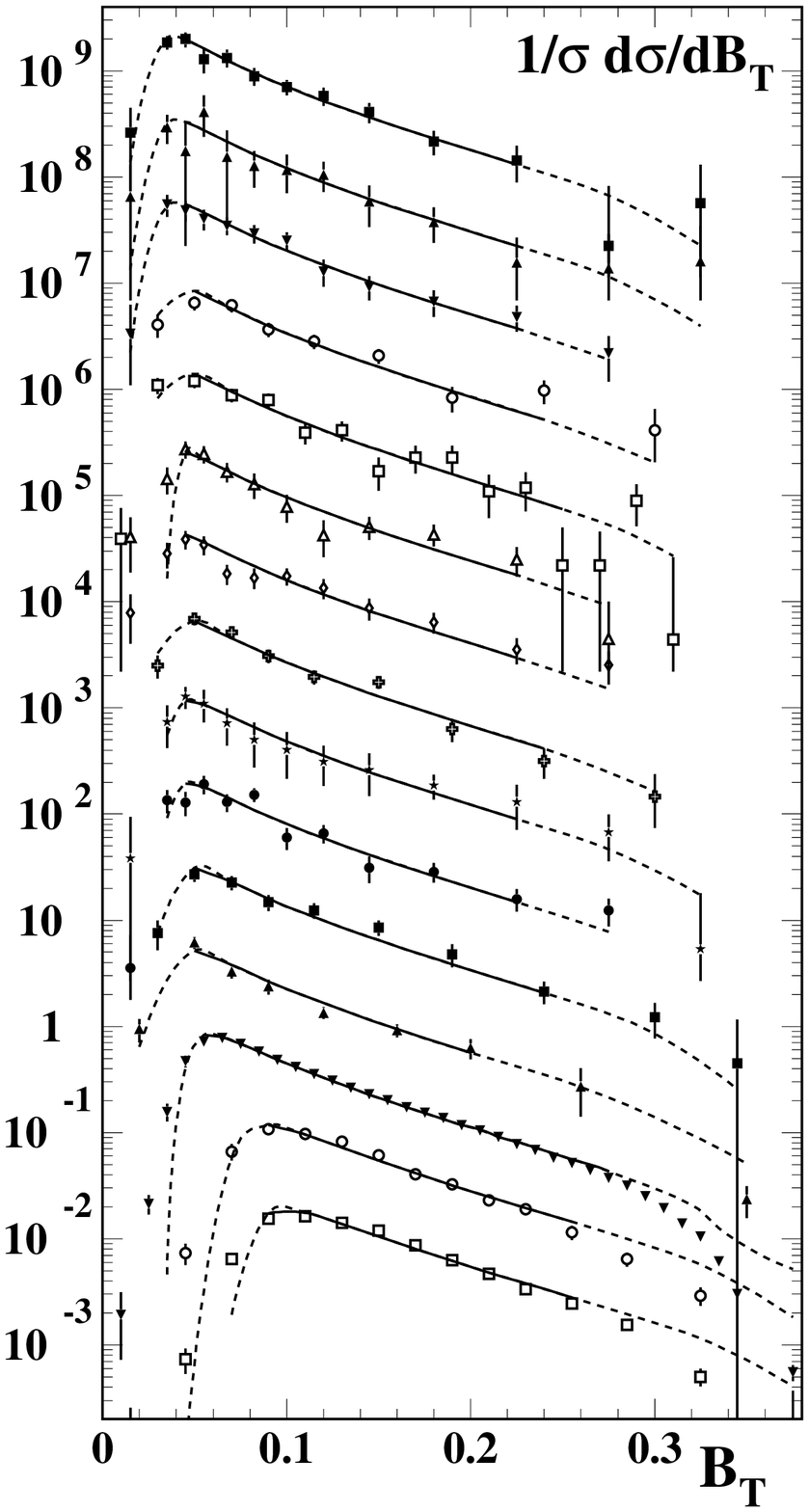,width=64mm}
    \hspace*{-15.35mm}\epsfig{file=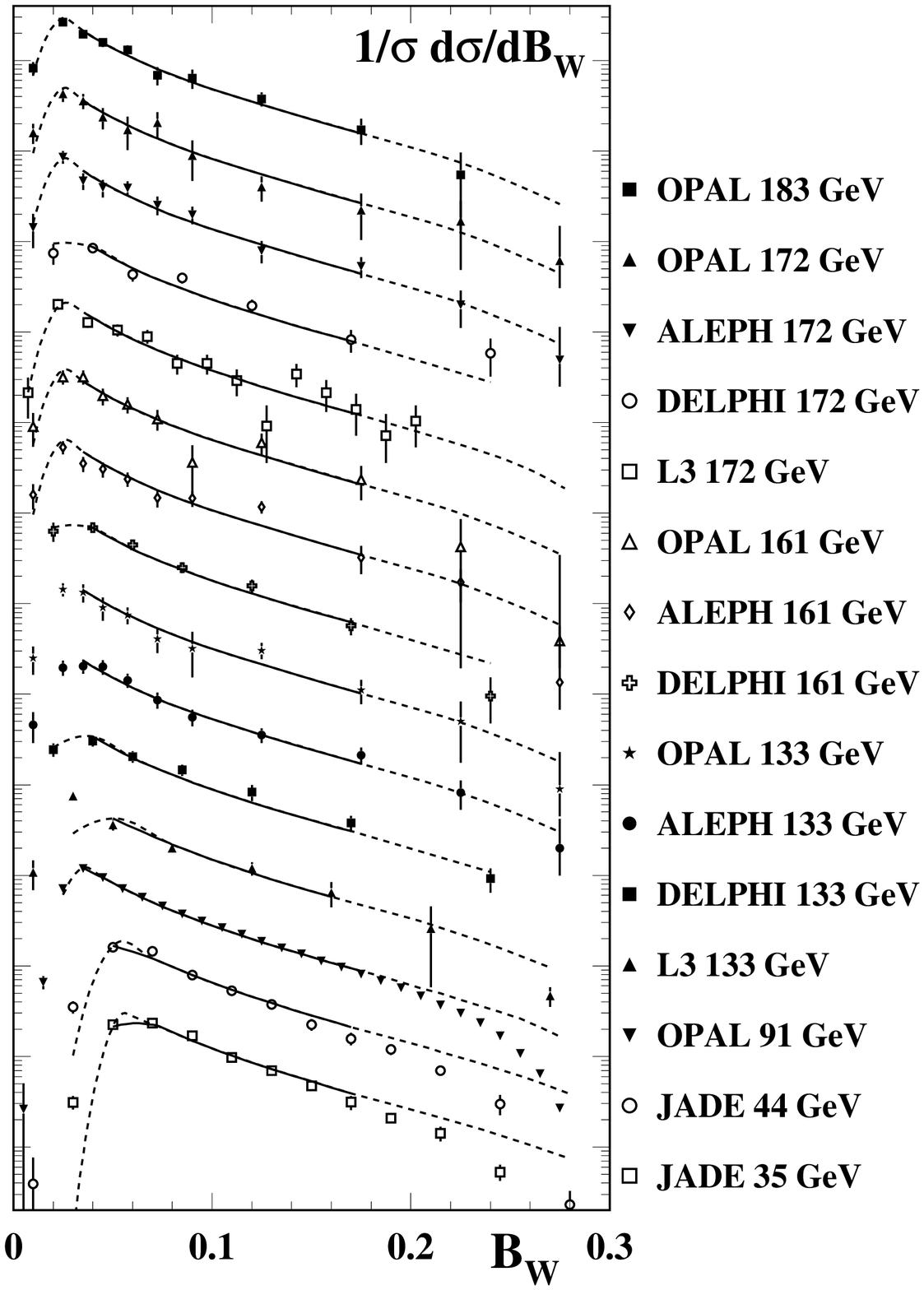,width=64mm}}
  \vspace*{-13mm}
\caption{\label{fig-global_fits} \small
  Simultaneous global fits of ${\cal O}(\alpha_s^2)$+NLLA theory +
  power corrections (lines) to the hadron level distributions of $C$
  parameter (left), total and wide jet broadening, $B_T$ and $B_W$
  (right), measured at $\protect\sqrt{s}$ = 35 to 183 GeV (points).
  The fit ranges are indicated by the solid lines.  The distributions
  shown are scaled.}  \vspace*{-3mm}
\end{figure*}

\subsection{Global Fits to PETRA and LEP data}
The universality of the power corrections is tested using differential
data distributions at hadron level measured with JADE at $\sqrt{s}$ =
35 and 44~GeV \cite{bib-fernandez,bib-fernandez-2}, and also using
data from various LEP experiments up to $\sqrt{s}$ =
183~GeV~\cite{bib-data}.  Both perturbative QCD (${\cal
  O}(\alpha_s^2)$+NLLA) and power corrections are compared by a global
simultaneous fit of $\alpha_s$, $\alpha_0$ and, in the case of the jet
broadenings, additionally $\alpha_0^{\prime}$ to the data.

Fig. \ref{fig-global_fits} shows the results of the fits (solid lines)
to the data (points).  Unlike in a similar analysis
in~\cite{bib-dokshitzer-2}, each energy has its individual fit range
including the maximum peaks in the 2-jet region. The fitted
distributions are in good agreement with the data within the fit
ranges.

In the case of the jet broadenings, a strong correlation between
$\alpha_0$ and $\alpha_0^{\prime}$ is observed, presumably because the
pure 1/$Q$ term $\xi{\cal P}+{\cal P^{\prime}}$ containing
$\alpha_0^{\prime}$ was found to be small and fairly energy
independent above $\sqrt{s}$ = 35~GeV.  A reanalysis of more data at
lower energies is needed to resolve both $\alpha_0$ and
$\alpha_0^{\prime}$.  For now, this contribution and, hence, the
dependence on $\alpha_0^{\prime}$ was neglected.

\begin{figure}[!t]
\vspace*{-5mm}
\centerline{\epsfig{file=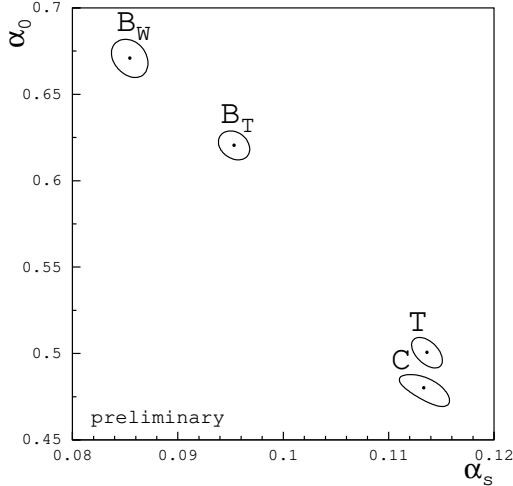,width=.42\textwidth}}
\vspace*{-12mm}
\caption{ \label{fig-con} \small
  Results for $\alpha_s$ and $\alpha_0$ from fits to event shape
  distributions, shown as 95 \% CL contours.}  \vspace*{-5mm}
\end{figure}

The corresponding fit results for $\alpha_s$($M_{Z^0}$) and
$\alpha_0$(2 GeV) are presented in table~\ref{tab-fit-results}, and in
Fig.~\ref{fig-con} with 95\% confidence level contours.  In the case of
$C$ parameter and --- for comparison --- thrust $T$, the results for
$\alpha_s$ are compatible with the world average value of
0.118~\cite{bib-bethke}.  Furthermore, there is a fair agreement of
the corresponding $\alpha_0$ values.  However, $\alpha_s$ from the
fits to the jet broadenings
\footnote{The 3-parameter fit of the full prediction for $B_T$ yields
  $\alpha_s$ = 0.0954 $\pm$ 0.0017, $\alpha_0$ = 0.710 $\pm$ 0.081,
  $\alpha_0^{\prime}$ = -0.707 $\pm$ 0.166 and a correlation between
  $\alpha_0$ and $\alpha_0^{\prime}$ of +98 \%. }
is far off the world average. Here, the universality of the effective
coupling moment is not observed.

\begin{figure}[!t]
  \vspace*{-7mm}
  \centerline{\hspace*{-6mm}\epsfig{file=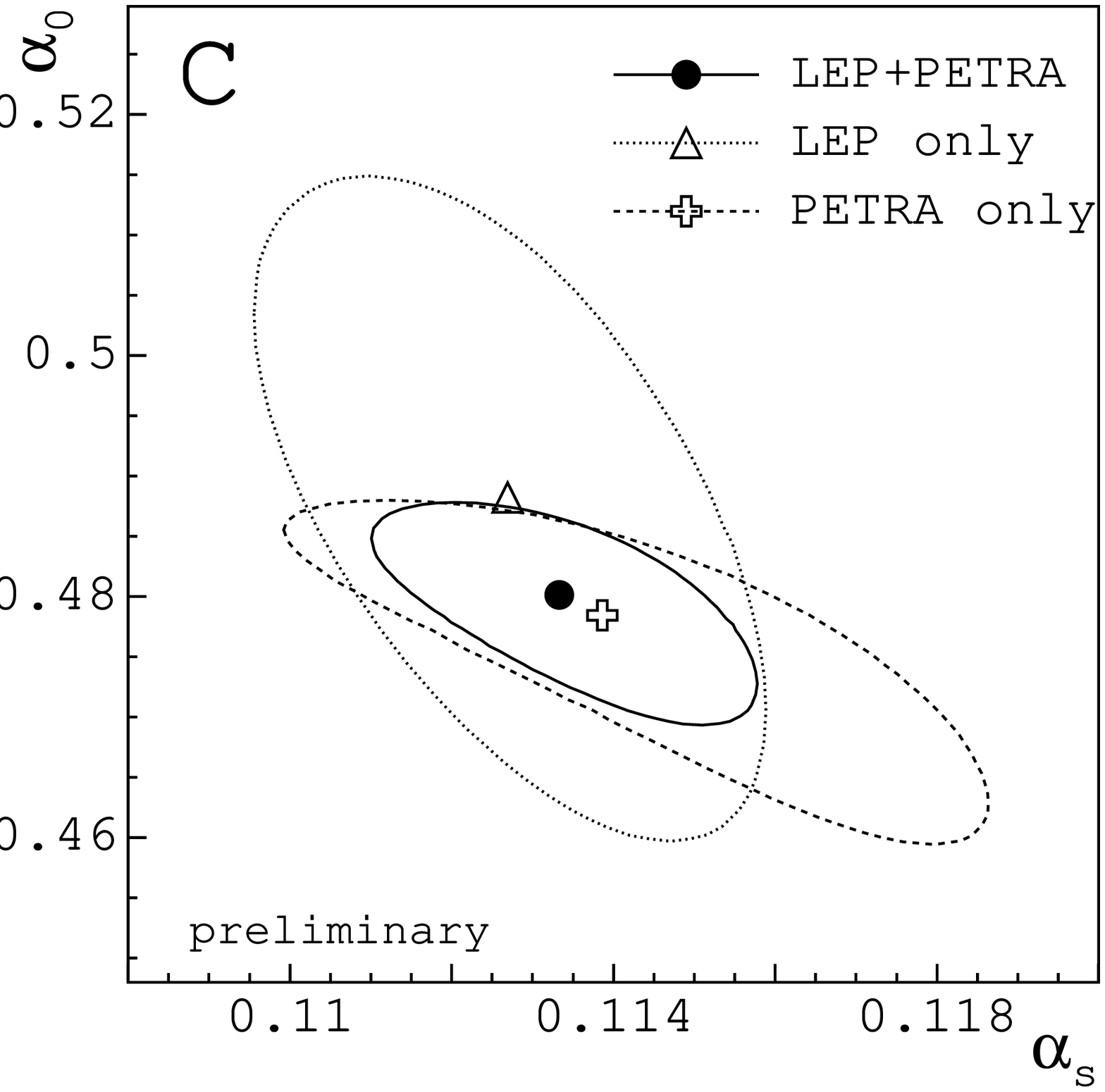,width=.31\textwidth}
    \hspace*{-6mm}
    \epsfig{file=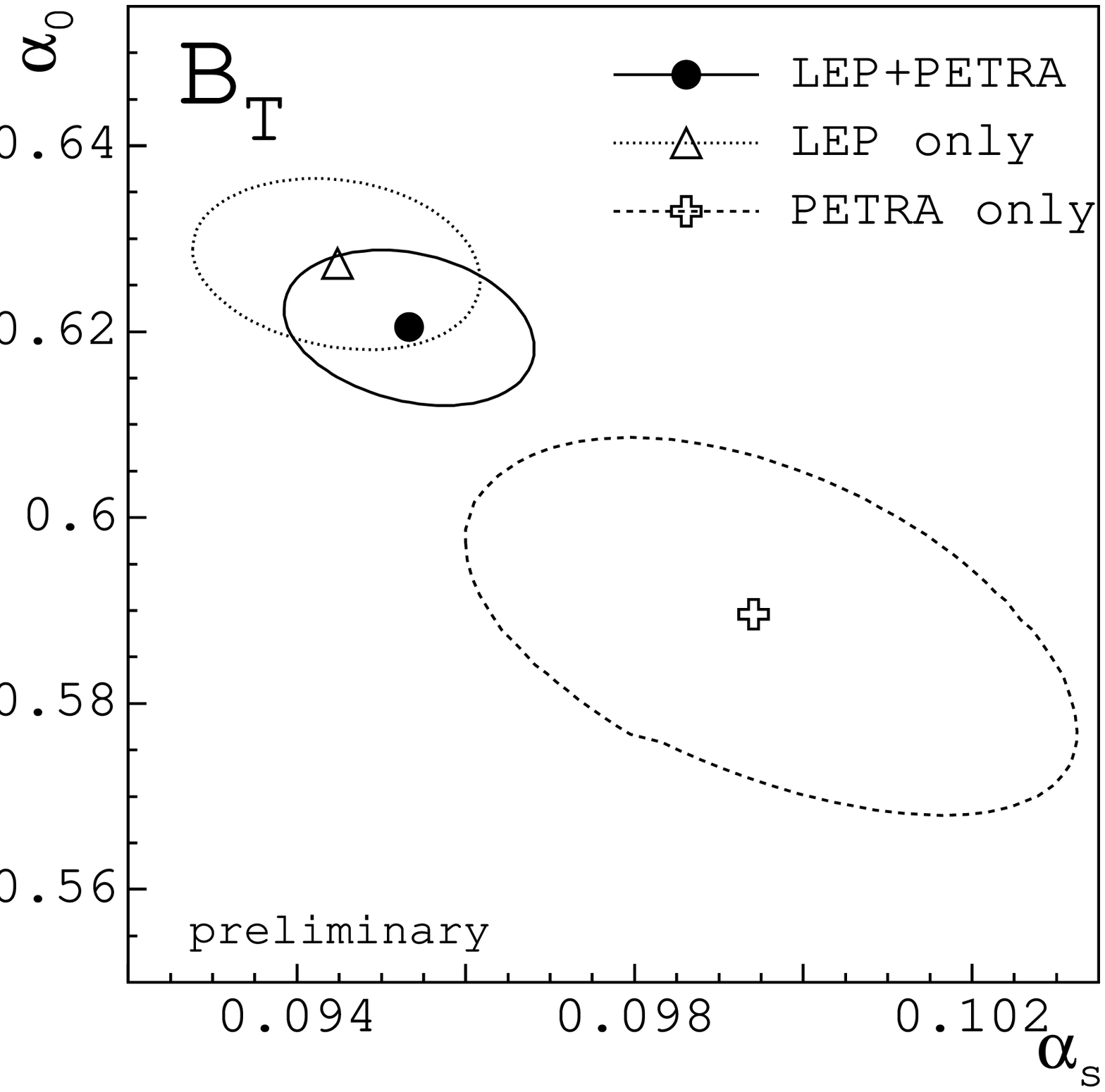,width=.31\textwidth}}
  \vspace*{-12mm}
\caption{ \label{fig-con_low+high}\small
  95 \% CL contours for $\alpha_s$ and $\alpha_0$ from fits to $C$
  and $B_T$ using PETRA only (dashed line), LEP only (dotted line) and
  combined (solid line) data.}  \vspace*{-7mm}
\end{figure}

The presence of possible 1/$Q^2$ contributions can be tested by
considering LEP data and JADE data separately in the analysis.
Fig.~\ref{fig-con_low+high} demonstrates that, in the case of the $C$
parameter, the JADE data reduce significantly the statistical
uncertainties of the fit. In the case of $B_T$, a minor indication for
1/$Q^2$ terms emerges.

\begin{table}[!b]
\vspace*{-10mm}
\begin{center}
\caption{\label{tab-fit-results}\small Preliminary results for $\alpha_s$ and $\alpha_0$.
  The errors stated are the statistical errors of the fit.}
\renewcommand{\arraystretch}{1.02}
\begin{tabular}{|c||r@{$\pm$}l|r@{$\pm$}l|c|}
\hline
  Obs.   &    \multicolumn{2}{c|}{$\alpha_s(M_{Z^0})$}    &    \multicolumn{2}{c|}{$\alpha_0$(2~GeV)} & $\chi^2$/dof \\
\hline\hline
$T$    & $ 0.1136$ & $ 0.0015 $  & $ 0.501$ & $ 0.009 $ & $ 248 / 226 $  \\
$C$    & $ 0.1128$ & $ 0.0022 $  & $ 0.482$ & $ 0.008 $ & $ 107 / 124 $  \\
$B_T$  & $ 0.0953$ & $ 0.0015 $  & $ 0.621$ & $ 0.008 $ & $ 159 / 141 $  \\
$B_W$  & $ 0.0854$ & $ 0.0017 $  & $ 0.671$ & $ 0.011 $ & $ 112 / 108 $  \\
\hline
\end{tabular}
\end{center}
\vspace*{-4mm}
\end{table}

A comparison of the fits based on distributions corrected for
hadronization as described in section \ref{sec-as_update} and the
hadron level fits using the power correction ansatz points out that
the transformation from the perturbative to the non-perturbative level
is not sufficiently described by a simple shift. Clearly, an
additional {\em squeeze} of the distributions is needed, i.e. power
corrections additionally depending on the value of the observable.  In
the present fits, $\alpha_s$ appears as the only parameter inducing a
squeeze. Thus, a combination of rather large values for $\alpha_0$ and
extremely small values for $\alpha_s$ is needed to fit the hadron
level distributions. This mechanism explains the observed
anticorrelation between $\alpha_s$ and $\alpha_0$ as seen in
Fig.\ref{fig-con}.  However, a shift is sufficient to describe the
impact of hadronization on the {\em means} of the event shape
distributions~\cite{bib-fernandez-2,bib-biebel}.

\section{SUMMARY}
Combined resummed and fixed order QCD provides consistent
determinations of $\alpha_s$ at PETRA energies. Recently presented
calculations for the $C$ parameter, $B_T$ and $B_W$ fit previously
published data at $\sqrt{s}$ = 35 and 44~GeV, yielding
$\alpha_s(M_{Z^0})$ = 0.123 $^{+0.008}_{-0.005}$ and 0.123
$^{+0.008}_{-0.006}$, respectively, in good agreement with the world
average value~\cite{bib-bethke}.

Power corrections to differential event shape distributions are tested
using hadron level data between $\sqrt{s}$ = 35 and 183~GeV.  Reliable
fits are obtained in the case of the $C$ parameter and the thrust,
where non-perturbative effects are approximately described by the
predicted shift of the perturbative spectra. This does not apply to
the case of $B_T$ and $B_W$, yielding $\alpha_s$ results incompatible
with other measurements at the $Z^0$ pole. For the present
calculations, the universality of the effective coupling moment
$\alpha_0$ at the level of 20\% is only verified for $C$ parameter and
thrust.

\end{document}